\documentclass[twocolumn,preprintnumbers, sort&compress, floatfix]{revtex4}
\usepackage{amsmath}
\usepackage{amsfonts}
\usepackage{amssymb}
\usepackage{graphicx}
\usepackage{bm}
\def\be{\begin{equation}}
\def\ee{\end{equation}}
\def\bi{\bibitem}

\begin{document}

\title{Unified cosmology with Non-minimally coupled scalar-tensor theory of gravity.}
\author{‪Behzad Tajahmad}
\email{behzadtajahmad@yahoo.com}
\affiliation{Faculty of Physics, University of Tabriz, Tabriz, Iran}
\author{Abhik Kumar Sanyal}
\email{sanyal_ak@yahoo.com}
\affiliation {Dept.of Physics, Jangipur College, Murshidabad, India - 742213}
\begin{abstract}

\noindent Unlike Noether symmetry, a metric independent general conserved current exits for non-minimally coupled scalar-tensor theory of gravity, if the trace of the energy momentum tensor vanishes. Thus, in the context of cosmology, a symmetry exists both in the early vacuum and radiation dominated era. For slow roll, symmetry is sacrificed, but at the end of early inflation, such a symmetry leads to a Friedmann-like radiation era. Late time cosmic acceleration in the matter dominated era is realized in the absence of symmetry, in view of the same decayed and red-shifted scalar field. Thus, unification of early inflation with late time cosmic acceleration with a single scalar field, may be realized.
\end{abstract}

\maketitle

\section{\bf{Introduction}}

A smooth luminosity-distance versus redshift curve of the distant SN1a Supernovae results in apparent dimming of the Supernovae than usual \cite{sn1, sn2}. These observations require present accelerated expansion of the universe. A host of dark energy models and their alternatives, viz., modified theories of gravity, exist in the literature, which explain the above phenomena. It is important to mention that after initial cosmic inflation, the universe should enter radiation dominated era. At this epoch, the cosmic scale factor must behave like Friedmann-Lamaitre standard model solution ($a\propto \sqrt t$). Again, after decoupling of CMBR photons and at the advent of pure matter dominated era ($z_{dec}\approx 1100$), the scale factor should behave like Friedmann-Lamaitre standard model solution, viz. ($a \propto t^{2\over 3}$). These are required to match other observational constraints like standard big-bang-nucleosynthesis (BBN), formation of cosmic microwave background radiation (CMBR), the epoch of matter radiation equality ($z_{eq} \approx 3200$), the observed epoch of decoupling ($z_{dec} \approx 1100$) and the beginning of matter dominated era, structure formation etc \cite{cmbr}. Analysis based on observational data also suggests that only recently, around redshift $z \approx 1$, the universe has entered accelerating phase, and the present value of the state parameter is $-0.5 \le\omega_0\le -1.5$. A viable cosmological model must accommodate all these features. Although $F(R)$ theory of gravity and its extended versions claim to have unified early inflation with late time cosmic acceleration, but this requires scalar-tensor equivalence at both ends \cite{odin, capo, mod}, which might be misleading, since physical equivalence has been questioned over decades \cite{1, 2, 3, 4, 5, 6, 7, 8, 9, 10, 11, 12, 13, 14, 15}. An attempt of unification has been made with non-minimally coupled dark energy model for the first time by Faraoni \cite{farao}. This motivates to explore the essence of non-minimally coupled models in further detail, in the context of such unification. \\

Over decades, Noether symmetry has been found to play important roles in explaining cosmic evolution \cite{n1, n2, n3, n4, n5, n6, n7, n8, n9}. However, for non-minimally coupled scalar-tensor theory of gravity, a metric independent conserved current is admissible directly from the field equations in general, which is not realized from Noether symmetry \cite{a1, a2, a3, a4}. Such a conserved current has been found to play a very important role to generate solutions, even for higher order theory of gravity \cite{a2, a3, a4}. In view of such non-Noether symmetry, here we show that the same non-minimally coupled scalar field can possibly drive the inflation at the very early stage of cosmic evolution, and although decays (slowly) in the process, explains late-time cosmological behaviour with extremely good precession, as well.

\section{The model and the symmetry}

Non-minimal coupling is unavoidable in a quantum theory of the scalar field $\phi$. Since such coupling is generated by quantum corrections, even if it is primarily absent in the classical action. Particularly, it is required by the renormalization properties of the theory in curved space-time background. Further, chaotic inflation with self interaction quartic potential ($V_0\phi^4$) proposed by Linde \cite{cinfl1} is disfavoured, since the spectral index of density perturbation ($n_s$) and the scalar to tensor ratio ($r$) do not agree with the constraints, viz. ($0.96 < n_s < 0.984$) and ($r < 0.14$) has been put up by recent Planck data \cite{Planck}. However, much before arrival of these data, Fakir and Unruh proposed an improvement on cosmological chaotic inflation taking nonminimal coupling into account \cite{improved}. We therefore start with the following action corresponding to non-minimally coupled scalar-tensor theory of gravity,
\be \label{action}\begin{split}A = &\int \Bigg[f(\phi) R - {\omega(\phi)\over \phi}\phi_{,\mu}\phi^{,\mu} - V(\phi) \\&- {1\over 2}\chi_{,\mu}\chi^{,\mu} - U(\chi)\Bigg]\sqrt{-g} d^4x\end{split}\ee
where, $f(\phi)$ and $\omega(\phi)$ are the non-minimal coupling parameter and the Brans-Dicke coupling parameter respectively, while $\phi$ and $\chi$ are two independent scalar fields having potentials $V(\phi)$ and $U(\chi)$ respectively. Corresponding field equations are

\be\label{fe1}\begin{split}& f(R_{\mu\nu} - {1\over 2} g_{\mu\nu} R) + \Box f g_{\mu\nu} - f_{;\mu;\nu}\\& -{\omega\over\phi}\phi_{,\mu}\phi_{,\nu}+{1\over 2}g_{\mu\nu} \left({\omega\over\phi}\phi_{,\mu}\phi_{,\nu}+V(\phi)\right) = {1\over 2}T_{\mu\nu}\end{split}\ee
\be \label{fe2}Rf' +2{\omega\over\phi}\Box\phi +\left({\omega'\over\phi}-{\omega\over\phi^2}\right)\phi^{,\mu}\phi_{,\mu}-V'(\phi) = 0.\ee
\be \label{fe3} \Box \chi - {dU\over d\chi} = 0.\ee
In the above, prime denotes derivative with respect to `$\phi$' and `$\Box$' stands for the D'Alembertian operator. Now taking the trace of equation (\ref{fe1})
\be R f -3\Box f-{\omega\over\phi}\phi^{,\mu}\phi_{,\mu} -2 V +{T\over 2} = 0.\ee
and eliminating the Ricci scalar ($R$) between equations (3) and (5), the above field equations may be cast in the form \cite{1}

\be \label{meq}\begin{split}&\left(3f'^2 + {2\omega f\over \phi}\right)^{1/2}  \left[\left(3f'^2 + {2\omega f\over \phi}\right)^{1/2}\phi^{;\mu}\right]_{;\mu}\\& - f^3 \left({V\over f^2}\right)'= {f'\over 2}\left[{\chi^{,\mu}\chi_{,\mu} + 4U(\chi)}\right]= {f'\over 2}{T^\mu}_\mu \end{split}\ee
In view of equation (\ref{meq}), a conserved current $J^\mu$ therefore exists where,

\be\label{cc} J^\mu_{;\mu} = \left[\left(3f'^2 + {2\omega f\over \phi}\right)^{1/2}\phi^{;\mu}\right]_{;\mu} = 0.\ee
for trace-less matter field, provided

\be \label{cond}V(\phi) \propto f(\phi)^2.\ee
It is important to mention that, while Noether symmetry may only be explored for finite degrees of freedom, the conserved current (\ref{cc}) is realized due to the presence of a general in-built symmetry of non-minimally coupled scalar-tensor theory of gravity. It is therefore also important to study the behaviour of such a conserved current (\ref{cc}) in different context. Here, we are particularly interested to study its behaviour in cosmological context. The motivation is to check how far it can accommodate present cosmological observations. In the spatially homogeneous and isotropic Robertson-Walker line element

\be \label{rw}ds^2 = - dt^2 + a^2(t) \left[\frac{dr^2}{1-kr^2} + r^2 (d\theta^2 + sin^2 \theta d\phi^2)\right]\ee
only three of the field equations (\ref{fe1}) through (\ref{fe3}) are independent, while there are seven [$a(t), \phi(t), f(\phi), \omega(\phi), V(\phi), \chi(t), U(\chi)$] field variables. Therefore, instead of choosing four of these field variables by hand, it is always instructive to use the conserved current (\ref{cc}), which now reads
\be\label{ccrw} \sqrt{\left(3f'^2 + {2\omega f\over \phi}\right)}a^3\dot\phi = C_1\ee
under the condition presented in equation (\ref{cond}). In the above, $C_1$ is the integration constant. Now, the additional scalar field $\chi$ may be treated as barotopic fluid inclusive of dark matter under the choice

\be \rho = {1\over 2}\dot\chi^2 + U(\chi),~~p = {1\over 2}\dot\chi^2 - U(\chi)\ee
where, $\rho$ and $p$ are the matter energy density and the pressure of the barotropic fluid under consideration respectively. Field equation (\ref{fe3}) is then simply the Bianchi identity

\be\label{bi} \dot\rho+3H({\rho+p}) = 0,\ee
where, $H = {\dot a\over a}$, is the Hubble parameter. It is important to note that, the energy-momentum tensor of the matter field is trace-less ($T = \rho-3p = 0$) both in the vacuum dominated ($\rho = p = 0$), and in the radiation dominated ($p = {1\over 3}\rho$) era. Thus, the conserved current (\ref{ccrw}) exists in these regime under the condition (\ref{cond}). However, in matter dominated era, the content of the universe behaves as pressure-less dust, $p = 0$, and so the trace of the energy-momentum tensor doesn't vanish. Hence, symmetry is broken in the matter dominated era. Now, to obtain exact solution of the field equations, we need two additional assumptions. Let us therefore make the following choice on $f(\phi)$, which automatically fixes the forms of $V(\phi)$ as,
\be\label{asmp1} f(\phi) = \phi^2 \Longrightarrow V(\phi) = V_0\phi^4.\ee
Further, if we choose
\be \label{asmp2} 3f'^2 + {2\omega f\over \phi}=\omega_0^2\ee
where $\omega_0$ is a constant, then the conserved current (\ref{ccrw}) reduces to
\be\label{cons} a^3\dot \phi = {C_1\over \omega_0}=C,\ee
$C$ being yet another constant. In the process, the form of and $\omega(\phi)$ is also fixed as
\be\label{omeg} \omega(\phi) = {\omega_0^2 - 12\phi^2\over 2\phi}.\ee
It is interesting to note that $a^3\dot\phi$ is a Noether conserved current, provided $\phi$ is cyclic. However, despite the presence of tight coupling of gravity with the scalar field $\phi$ through the coupling parameter $f(\phi)$, non-trivial Brans-Dicke parameter $\omega(\phi)$ and the quartic potential $V(\phi)$, the same conserved current has been found to exist. Thus, the symmetry under consideration is non-Noether type, and of-course is more general. At this stage, the functions $f(\phi)$, $V(\phi)$ and $\omega(\phi)$ have been expressed explicitly as function of $\phi$, and the field equations therefore read
\be\begin{split}
\label{fa}&\left(2{\ddot a\over a}+{\dot a^2\over a^2}+{k\over a^2}\right)+2\left({\ddot\phi\over \phi}+{\dot\phi^2\over \phi^2}+2{\dot a\over a}{\dot\phi\over\phi}\right)\\&+\left(\omega_0^2-12\phi^2\over 4\phi^2\right){\dot\phi^2\over\phi^2}-{1\over 2}V_0\phi^2  =  -{p\over 2\phi^2}.\end{split}\ee
\be\begin{split}\label{fphi} 6\left({\ddot a\over a}+{\dot a^2\over a^2}+{k\over a^2}\right)&-{\omega_0^2-12\phi^2\over 2\phi^2}\left({\ddot\phi\over \phi}+3{\dot a\over a}{\dot\phi\over\phi}\right)\\&+{\omega_0^2\over 2\phi^2} {\dot\phi^2\over\phi^2} - 2 V_0\phi^2 =  0.\end{split}\ee
\be\begin{split}\label{fh} 3\left({\dot a^2\over a^2}+{k\over a^2}\right) + 6{\dot a\over a}{\dot\phi\over\phi} - {\omega_0^2-12\phi^2\over 4\phi^2}{\dot \phi^2\over\phi^2} - {1\over 2}V_0\phi^2 = {\rho\over 2\phi^2}.\end{split}\ee
Equations (\ref{fa}) and (\ref{fphi}) are the $a~\mathrm{and } ~\phi$ variation equations respectively, while equation (\ref{fh}) is the $(^0_0)$ equation of Einstein, and is also known as the Hamilton constraint equation, when expressed in terms of phase-space variables.

\section{Three stages of cosmic evolution}

\subsection{Early universe}

In the vacuum dominated era $p = 0= \rho$, and so the action \eqref{action} may be expressed in the form
\be A = \int \Big[f(\phi) R - {K(\phi)\over 2}\phi_{,\mu}\phi^{,\mu} - V(\phi)\Big]\sqrt{-g} d^4x,\ee
Inflation with such non-minimal coupling is undergoing serious investigation over decades \cite{farao, s1, s2, s3, s4, s5, s6, s7, s8, s9, s10}. In the Einstein's frame (under conformal transformation $g_{\mu\nu}^E = f(\phi) g_{\mu\nu}$), the above action may be cast as \cite{simon}

\be A = \int \Big[R_E - {1\over 2}(\partial_E \sigma)^2 - V_E(\sigma(\phi))\Big]\sqrt{-g_E} d^4x,\ee
where,
\be V_E = {V(\phi)\over f^2(\phi)}, ~~\left({d\sigma\over d\phi}\right)^{2} = {K(\phi)\over f(\phi)} + 3 {f'(\phi)^2\over f(\phi)^2}\ee

In the non-minimal theory, the flat section of the potential $V(\phi)$, responsible for slow rollover, is usually distorted. Generalizing the form of non-minimal coupling by an arbitrary function $f(\phi)$, Park and Yamaguchi \cite{PY} could show that the flat potential is still obtainable when $V_E$ is asymptotically constant. Let us therefore relax the symmetry and choose the potential in the form
\be V = V_0\phi^4 - B\phi^2,\;\mathrm{so}\;\mathrm{that},\;V_E = V_0 - {B\over \phi^2}.\ee
Therefore, initially when $\phi \gg \sqrt {B\over V_0}$, the second term may be neglected, and $V_E = V_0$, becomes flat to admit slow roll. Now the slow roll parameters are
\be \epsilon = \left({V_E'\over V_E}\right)^{2}\left(d\sigma\over d\phi\right)^{-2} = {4B^2\phi^2\over \omega_0^2(V_0\phi^2-B)^2}\ee
\be \begin{split}\eta &= 2\left[{V_E''\over V_E}\left(d\sigma\over d\phi\right)^{-2}-{V_E'\over V_E}\left(d\sigma\over d\phi\right)^{-3}{d^2\sigma\over d\phi^2}\right]\\&={4B\phi^2\over \omega_0^2(V_0\phi^2 -B)}\end{split}\ee
The number of e-foldings of slow-roll inflation is given by an integral over $\phi$ as,
\be N_e(\phi) = {1\over 2}\int_{\phi_{e}}^{\phi_{b}} {d\phi\over \sqrt{\epsilon(\phi)}}{d\sigma\over d\phi} = {\omega_0^2\over 4 B}\left({B\over 2\phi^2}+V_0\ln \phi\right)_{\phi_e}^{\phi_b}\ee
where, $\phi_b$ is the initial value $\phi$ when inflation starts, and $\phi_e$ is its final value, when inflation ends, i.e. at $\epsilon \approx 1$. Now, under the choice, $V_0 = 1.0\times10^{-13}$, $\omega_0 = 3.26\times 10^{-3}$, $B = 0.79\times 10^{-21}$ and $\phi_b = 1.26\times 10^{-4}$, we obtain $\epsilon = 0.0059,~ \eta = 0.059$. Therefore, the spectral index ($n_s$) and the scalar to tensor ratio ($r$) take the following values,
\be  n_s = 1 -6\epsilon + 2\eta = 0.976;\;\;\;r = 16\epsilon = 0.094. \ee
which agree fairly well with recently released Planck TT + low P + lensing data, which puts even tight constraints on these parameters viz., $n_s < .984$ and $r < 0.14$ \cite{Planck} as already mentioned. The end of inflation $\epsilon =1$, occurs at the value of the scalar $\phi_e = 0.9134 \times 10^{-4}$. This results in $N_e = 33$$e$-folding, which although appears to be a little less than usual, viz. $50 < N < 60$, it doesn't matter, since the precise number of e-folding at which the present Hubble scale $k = a_0H_0$ is equal to the Hubble scale during inflation is model dependent and the objective is to get enough e-folding such that inflation ends, leading to a flat universe. The same above set of data with a slightly different value of $V_0 = 1.1\times 10^{-13}$ yield, $\epsilon = .0041$, $\eta = .0049$, $n_s = 98.4$ and $r = .065$. Inflation in this case ends at $\phi_e = 0.86 \times 10^{-4}$, which gives $N_e = 45$$e$-folding. This appears to be even better, although $n_s$ is at its limit. It might appear that $\phi_b$ is small enough, it again doesn't matter, since results are consistent. However, it is to be mentioned that the value of $\phi_b$ has been chosen in accordance with the values of $V_0$ and $\omega_0$, which give excellent fit with observed data at late-time cosmological evolution, as we shall see later.

\subsection{Radiation era}
The most compelling feature of non-minimally coupled scalar-tensor theory of gravity is that, the scalar field $\phi$ decays via gravitational effects. This is possible because, coupling between the scalar and matter fields arises spontaneously when $\phi$ settles down to its vacuum expectation value $<\phi>$ and oscillates, ensuring  reheating of the universe \cite{reheat1, reheat2, reheat3}. In particular, it has been shown that reheating occurs in a broad class of $f(\phi)R$ models \cite{reheat4}. We therefore presently just assume that it may also be possible for the present model, and attempt to show the same in future. Now after graceful exit from inflationary regime, the universe enters radiation dominated era. But, in pure radiation era if the scale factor $a \propto \sqrt t$, i.e. if the universe evolves like standard Friedmann solution, then only standard Nucleosynthesis, and structure formation are realized. This is possible if somehow the second term in the potential vanishes at the end of inflation, which we demonstrate below. But, actually following the decay of the scalar field $\phi$, $V_0\phi^4$ and $B\phi^2$ are now of the same order of magnitude. So if one ignores the second term \footnote{There is one possibility, to neglect the term at both ends of $\phi$, and that is by choosing $B = k$ - the curvature parameter, then at the beginning $k > 0$ (say), ensuring positive curvature, while at the end of inflation, the universe becomes flat ($k = 0$), and the second term vanishes. However, we can't present any physical argument behind such a choice of potential, rather only rely on the fact that - it works. However, here it is not possible, since the value of $B$ is much smaller.}, or associates with the first (since both are small and of the order of $10^{-29}$), then one can restore symmetry in the radiation era. Since, in view of equations (\ref{fa}) and (\ref{fh}), one can construct yet another equation, viz,

\be\label{combination}\begin{split}&6\left({\ddot a\over a}+{\dot a^2\over a^2}+{k\over a^2}\right)+6\left({\ddot\phi\over \phi}+{\dot\phi^2\over\phi^2}+3{\dot a\over a}{\dot\phi\over\phi}\right)\\&+\left({\omega_0^2-12\phi^2\over 2\phi^2}\right){\dot\phi^2\over\phi^2} - 2 V_0\phi^2 = {1\over 2\phi^2}(\rho - 3p) = 0.\end{split}\ee
Further, differentiating the conserved current (\ref{cons}) twice and doing a little algebraic manipulation, one obtains

\be \label {H}3\left({\ddot a\over a}+{\dot a^2\over a^2} +{k\over a^2}\right)= -{\stackrel{...}\phi\over\dot\phi} + {5\over 3}{\ddot\phi^2\over\dot\phi^2}+3k\left({\dot\phi\over C}\right)^{2\over 3}.\ee
The last couple of equations (\ref{combination}) and (\ref{H}) may be combined together to eliminate the scale factor, resulting in

\be \label{scalar}\dot\phi\stackrel{...}\phi - {5\over 3}\ddot\phi^2 -{\omega_0^2\over 4}{\dot\phi^4\over\phi^4} - 3k\left({\dot\phi^4\over C}\right)^{2\over 3} + V_0\phi^2\dot\phi^2 = 0.\ee
The last equation (\ref{scalar}) admits a solution for the scalar field in the form
\be \phi = {\phi_0\over \sqrt{At -t_0}}\ee
where, $A$ and $\phi_0$ are constants of integration and the third constant has been absorbed. The scale factor therefore evolves as,

\be a = a_0 \sqrt{At - t_0}\ee
Note that, in view of the above solutions $a(t) \phi (t) = \mathrm{const}$. The interesting fact is that despite tight coupling with the scalar field, radiation era admits Friedmann-like solution. Thus, nucleosynthesis, era of matter-radiation equality, structure formation and decoupling epoch remain unaffected.\\

\subsection{Matter dominated era}

As already noticed, during matter dominated era, the matter content of the universe behaves like pressure-less dust, $p = 0$, and therefore, the trace of the energy-momentum tensor doesn't vanish ($T \ne 0$), as a result, the cherished symmetry failed to exist. Hence, there is no conserved current (\ref{ccrw}). It is therefore impossible to solve the set of field equations (\ref{fe1}) through (\ref{fe3}) analytically. Nevertheless, since the forms of the coupling parameter, potential and the Brans-Dicke parameter $\big[f(\phi), V(\phi), \omega(\phi)\big]$ are known in view of the relations (\ref{asmp1}) and (\ref{omeg}), so we try for numerical solution in the flat space, $k = 0$. First of all note that Bianchi identity (\ref{bi}) leads to $\rho a^3 = \rho_0$, where, the constant $\rho_0$ is the matter content of the universe at present. Using the expression for the matter density and the chosen form (\ref{asmp2}) one can reduce relation (\ref{meq}) to the following form,

\be\label{matter} \omega_0^2 {d\over dt}(a^3\dot\phi) = -{\rho_0\over a^3} = -\rho.\ee
We plug-in the form of $\rho$ so obtained, on the right hand side of the $(^0_0)$ component of Einstein's equation (\ref{fe3}) and solve the set of equations (\ref{fe1}) and (\ref{fe3}) with Runge-Kutta $4_{th}$-order approach. Since, after decoupling of CMBR photons ($z = 1100$), pure matter dominated era started, so in the present approach, we set initial values of $a, \dot a, \phi, \dot \phi$ at $z = 1100$, for which we have chosen $t = 4,90,000~ \mathrm{years}$. We have also set the parametric values of $V_0$ and $\omega_0$ according to table - 1.\\

\begin{center} \textbf{Table - 1}\end{center}
\begin{tabular}{|c|c|c|}
  \hline
  Parameters & Values & Range\\
  \hline
  $V_0$ & $ 1.0\times 10^{-13}$ & $(0.97~\mathrm{to}~ 1.15)\times 10^{-13}$\\\hline
  $\omega_0$ & $3.26\times 10^{-3}$ & $(3.25~\mathrm{to}~ 3.27)\times 10^{-3}$\\\hline
  $a(z = 1100)$ & $9.08\times 10^{-4}$ & $(9.07 ~\mathrm{to} ~9.09)\times 10^{-4}$\\\hline
  $\phi(z = 1100)$  & $3.17\times 10^{-8}$ & $(3.16~\mathrm{to}~ 3.18)\times 10^{-8}$ \\\hline
  ${da\over dt}(z = 1100)$  &  $1.24\times 10^{-9}$ & $(1.235 ~\mathrm{to}~ 1.245)\times 10^{-9}$\\\hline
  ${d\phi\over dt}(z = 1100)$  &$6.48\times 10^{-14}$ & $(6.475~\mathrm{to}~6.485)\times 10^{-14}$\\
  \hline
\end{tabular}\\

\noindent
In the above set of parametric values, we have kept the same $V_0$ and $\omega_0$, that was required to drive the inflation. Also, we remember that the scalar field at the end of inflation was $\phi_e = 0.79 \times 10^{-4}$. But, at the advent of matter dominated era, we have chosen it few order of magnitude less. This is because, at the end of inflation, the scalar field oscillates and decays via gravitational effect and finally the rest has been redshifted according to the solution (31) almost for $10^5$ years, till photons decoupled. Other initial data have been chosen under trial and error, so that these might lead to late time acceleration having excellent fit with the presently available cosmological observations. We present several plots to demonstrate the results obtained in the present scheme. The scale factor versus proper-time plot (figure-1) gives the following qualitative behavior of the scale factor,
\be a \propto \sinh{t}^{2\over 3}.\ee
So at early stage of matter dominated era, the universe had undergone Freedman-like decelerated expansion ($a \propto t^{2\over 3}$), and accelerated expansion started at the late stage of cosmic evolution. The scale factor versus redshift plot (figure-2), confirms that the present value of the scale factor is exactly $1$. The evolution of the Hubble parameter is shown in Hubble parameter versus redshift graph (figure-3), which gives its present value, $H_0 = 7.15 \times 10^{-11} \mathrm{yr^{-1}}$. The age versus redshift plot (not presented) indicates that the present age of the universe is $t_0 = 13.86~ \mathrm{Gyr}$. Therefore $H_0 t_0 = 0.991$, which fits the observational data with high precision, and has been demonstrated in figure-4. Figure-5 represents deceleration parameter versus redshift plot. The top inset plot demonstrates that $q = 0.5$ till $z = 200$. It then falls very slowly and at around $z = 4$, it takes the value $q \approx 0.48$. Afterwards, it falls sharply and acceleration starts at $z = 0.75$, as depicted in the inset plot below. The present value of deceleration parameter is $q = -0.59$. The present value of the effective state parameter is therefore, $w_{eff0} = -0.73$. As usual, ignoring small variation of the prefactor, we consider that the CMBR temperature falls as $a^{-1}$. If we consider the CMBR temperature at decoupling to be $T_{dec} \approx 3000~\mathrm{K}$ \cite{fixen}, then the present value of it is, $T_0 = 2.7255 ~\mathrm{K}$. In figure-6, we have presented the time - temperature graph. In a nutshell, our findings are,

\begin{figure}
  \centering
  \includegraphics[width=2.5 in, height=2.2 in]{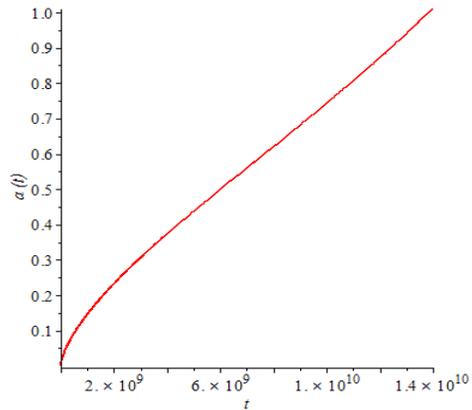}\\
  \caption{Qualitative behaviour of the Scale factor $a \propto \sinh^{2\over3}{t}$ is apparent in $a$ versus $t$ plot. Therefore at early epoch, the scale factor evolves as $t^{2\over 3}$, confirming early deceleration and late-time acceleration.}\label{1}
\end{figure}

\begin{figure}
  \centering
  \includegraphics[width=2.5 in, height=2.2 in]{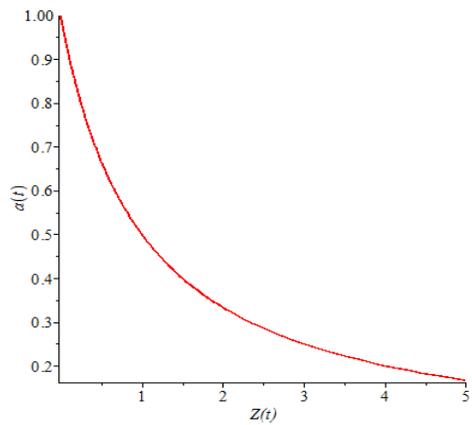}\\
  \caption{Scale factor $a$ versus redshift $z$ plot confirms that the present value of the scale factor is $a_0 = 1$.}\label{2}
\end{figure}

\begin{figure}
  \centering
  \includegraphics[width=2.5 in, height=2.2 in]{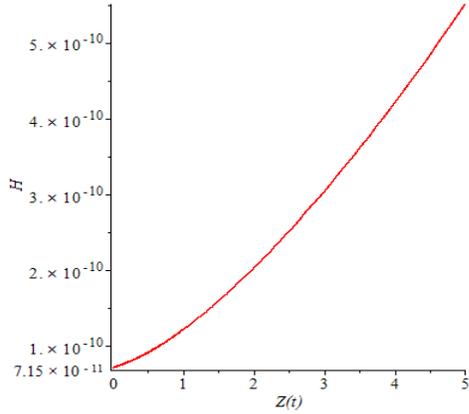}\\
  \caption{The Hubbble parameter $H$ versus redshift $z$ plot gives the present value of the Hubble parameter $H_0 = 7.15 \times 10^{-11} \mathrm{yr^{-1}}$. This value corresponds to $H_0 = 69.96 ~\mathrm{Km. s^{-1} Mpc^{-1}}$}\label{3}
\end{figure}

\begin{figure}
  \centering
  \includegraphics[width=2.5 in, height=2.2 in]{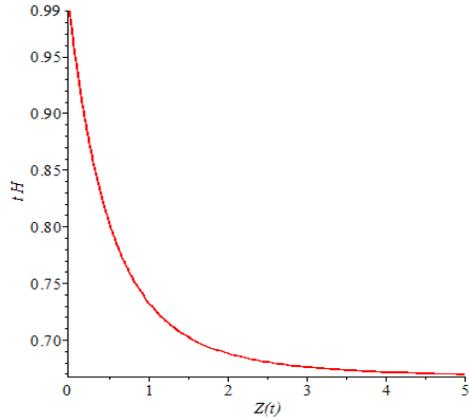}\\
  \caption{The $Ht$ versus $z$ plot shows present value is $H_0\times t_0 = 0.99$}\label{4}
\end{figure}

\begin{figure}
  \centering
  \includegraphics[width=3.4 in, height=3.2 in]{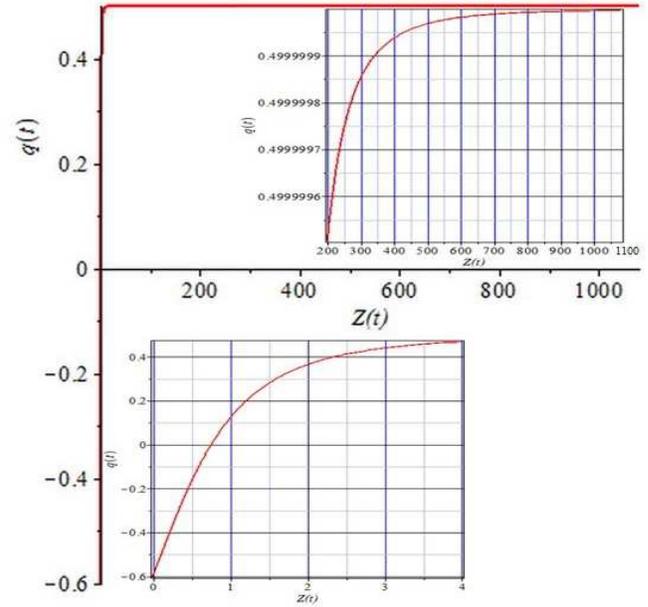}\\
  \caption{Deceleration parameter $q$ versus redshift $z$ plot indicates a long matter dominated era till $z \approx 4$. Accelerated expansion starts at $z = 0.75$. The present value of the deceleration parameter $q_0 = -0.59$.}\label{5}
\end{figure}

\noindent
1. Age of universe $t_0 = 13.86~ \mathrm{Gyr}$.\\
2. The universe undergoes Freedmann-like matter dominated era for quite a long time, before entering accelerated expansion epoch. Acceleration starts at the redshift value, $z = 0.75$ which is at about half the age of the universe, $t_{acceleration} = 7.20~ \mathrm{Gyr}$.\\
3. Present value of the scale factor, $a_0 = 1.00$. \\
4. Present value of the Hubble parameter, $H_0 = 7.15\times10^{-11}~ \mathrm{yr^{-1}}$, which is equivalent to $H_0 = 69.96~\mathrm{Km.s^{-1}Mpc^{-1}}$.\\
5. $H_0\times t_0 = 0.99$.\\
6. Present value of the deceleration parameter, $q _0 = - 0.59$.\\
7. The deceleration parameter, $q$ remains almost constant, $q \approx 0.5$, till $z = 4.0$, confirming a long Freedmann-like matter dominated era. \\
8. Present value of the state parameter, $\omega_{eff _0} = - 0.73$.\\
9. Present value of $\phi$ is $9.0\times 10^{-4}$.\\

\noindent
Further, from the relations

\be\label{dark} \rho_{\phi} = \phi^2\left[-6{\dot a\dot\phi\over a\phi}+{1\over 4}{\omega_0^2 - 12\phi^2\over\phi^4}\dot\phi^2 + {1\over2}V_0\phi^2\right]\ee
\be \label{roc}\rho_{c} = 3 {\dot a^2 \over a^2}\phi^2,\ee
we have found $\Omega_{\phi_0} = {\rho_{\phi_0}\over \rho_{c}} = 0.71 = 71\%$ and hence, $\Omega_{m0} = 0.29 = 29\%$. All these results have been found to remain unaltered within the specified range presented in the table - 1.\\

\begin{figure}
  \centering
  \includegraphics[width=2.5 in, height=2.2 in]{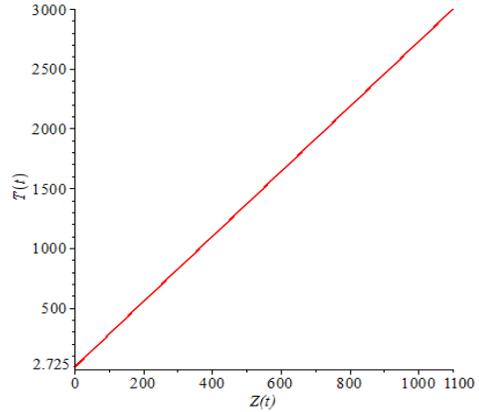}\\
  \caption{The CMBR temperature $T$ versus redshift $z$ plot shows presently $T_0 = 2.725$.}\label{6}
\end{figure}

\noindent
We have also studied the behaviour of the state-finder using the following relations

\be\label{sfr} r = q+2q^2 -{{\dot q}\over H};\;\;\;s = {r-1\over 3(q-1/2)}.\ee
Numerical analysis shows that the presently $\{r, s\} \approx \{1, 0\}$, presented in figure-7. Thus the correspondence of the present model with the standard $Λ\mathrm{CDM}$ universe model has also been established.\\

\begin{figure}
  \centering
  \includegraphics[width=2.5 in, height=2.2 in]{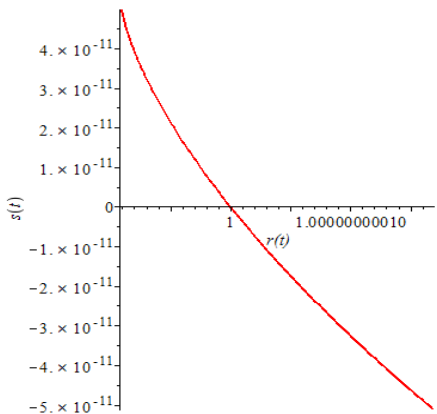}\\
  \caption{The state-finder $s$ versus $r$ plot represents perfect correspondence of the present model with the standard $Λ\mathrm{CDM}$ model}\label{7}
\end{figure}

\noindent
It is usually argued \cite{farao} that the self-coupling constant $V_0$ of the scalar field in the chaotic inflation potential $V = V_0 \phi^4$ is subject to the constraint $V_0 < 10^{-12}$ coming from the observational limits on the amplitude of fluctuations in the cosmic microwave background. This constraint makes the scenario uninteresting because the energy scale predicted by particle physics is much higher. We therefore relax the constraint on $V_0$ by and large and observe that for a totally different set of initial and the parametric values of $V_0,~\omega_0$, presented in table-2, the qualitative behaviour of cosmic evolution remains almost unaltered.

\begin{center} \textbf{Table - 2}\end{center}
\begin{tabular}{|c|c|c|}
  \hline
  Parameters & Values & Range\\
  \hline
  $V_0$ & $ 1 $ & $0.98~\mathrm{to}~ 1.13$\\\hline
  $\omega_0$ &$ 6.86\times 10^{-11}$ & $(6.85~\mathrm{to}~ 6.87)\times 10^{-11}$\\\hline
  $a(z = 1100)$ & $9.07\times 10^{-4}$ & $(9.06 ~\mathrm{to} ~9.08)\times 10^{-4}$\\\hline
  $\phi(z = 1100)$  & $0.1$ & $0.099~\mathrm{to}~ 0.101$ \\\hline
  ${da\over dt}(z = 1100)$  &  $1.19\times 10^{-9}$ & $(1.185 ~\mathrm{to}~ 1.195)\times 10^{-9}$\\\hline
  ${d\phi\over dt}(z = 1100)$  &$-4.08\times 10^{-7}$ & $- (4.075~\mathrm{to}~4.085)\times 10^{-7}$\\
  \hline
\end{tabular}\\

\noindent
In the above, we assume that at a redshift $z=1100$, the age of the universe was $t=510,000 ~\mathrm{yrs}$. The present values obtained in the process are enlisted underneath.\\

\noindent
1. Age of universe $t_0 = 14.26~ \mathrm{Gyr}$.\\
2. The universe undergoes Freedmann-like matter dominated era for quite a long time, before entering accelerated expansion epoch. Acceleration starts at the redshift value, $z = 0.78$ which is at about half the age of the universe, $t_{acceleration} = 7.20~ \mathrm{Gyr}$.\\
3. Present value of the scale factor, $a_0 = 1.00$. \\
4. Present value of the Hubble parameter, $H_0 = 7.08\times10^{-11}~ \mathrm{yr^{-1}}$, which is equivalent to $H_0 = 69.24~\mathrm{Km.s^{-1}Mpc^{-1}}$.\\
5. $H_0\times t_0 = 1.01$.\\
6. Present value of the deceleration parameter, $q _0 = - 0.62$.\\
7. The deceleration parameter, $q$ remains almost constant, $q \approx 0.5$, till $z = 5.76$, confirming a long Freedmann-like matter dominated era. \\
8. Present value of the state parameter, $\omega_{eff _0} = - 0.75$.\\
9. Present value of $\phi$ is $1.33\times 10^{-10}$.\\
10. Taking $T_{dec}=3004.41$, at a redshift $z = 1100$, the present value of CMBR temperature is $T_0 = 2.725$.\\
11. $\{r, s\} \approx \{1,0\}$.\\
12. Here again $\Omega_{\phi_0} = 71\%$ and so $\Omega_{m0} = 29\%$.\\
It is needless to present the plots, since as already mentioned, the qualitative behaviour remains unaltered. Thus, it is clear that the technique works fairly well, for a wide range of parametric and initial values. However, inflation has not been tested for these second set of initial values.

\section{Conclusion}

The same scalar, responsible for early inflation, resulting in the inflationary parameters ($r < 0.10,~ n_s \approx .98$ which are at par with recent Planck data), gives way to Friedmann-like solution ($a\propto \sqrt t$) in the radiation era, a Friedmann-like long matter dominated era ($q = 0.5$ till $z \approx 4$) and late-time accelerated expansion at around red-shift $z \approx 0.75$.The present value of the Hubble parameter ($H_0 \approx 69.96~\mathrm{Km. ~s^{-1}~Mpc^{-1}}, H_0t_0 =0.99$) are also at par with Planck's data. These results are definitely interesting. However, the claim that - a single scalar field might possibly solve the cosmological puzzle singlehandedly, can only be made after one can show that the field oscillates and reheat the universe sufficiently, which we pose to attempt in future.\\

This has been possible since a general conserved current is associated with non-minimally coupled scalar-tensor theory of gravity. One has to give up the conserved current at the very early universe, since the potential $V = V_0 \phi^4$ needs to be modified to $V(\phi) =  V_0 \phi^4 - B \phi^2$, for admitting slow-roll. When $\phi$ is large, second term may be neglected, resulting in a flat potential. Usually, the discussion with scalar field models ends up, after giving way to the radiation era. Now, to keep all the calculations (Baryogenesis, Nucleosynthesis, growth of perturbation  etc.) based on standard model unaltered, the scale factor in the radiation era must evolve like Friedmann-model ($a\propto \sqrt t$). This may be achieved, provided, one can restore the symmetry, which requires to neglect the second term yet again. In the present situation, the parameters and $\phi_b$ have been so chosen that, neglecting the second term in the potential don't create any problem. However, as we have suggested [52], this may be done, if one chooses $B = k$, where, $k$ - the curvature parameter vanishes making the universe flat at the end of Inflation. Although, we can't present any physical argument behind such choice, still this may be attempted in future.\\

The results remain almost unaltered over a wide range of initial and parametric values and it appears that one can tinker with these values to obtain even better result. However, at the end, let us mention that there is a shuttle important difference in the cosmic evolution arising out of the two sets of data presented in table 1 and 2 also. According to the first set of data, $\phi|_{z=1100} = 3.17\times 10^{-8}$, and its present value is $\phi|_{0} = 9.0\times 10^{-4}$. Therefore the scalar field as well as the potential increases. In the second set, $\phi|_{z=1100} = 0.1$, and its present value is $\phi|_{0} = 1.33\times 10^{-10}$. Therefore $\phi$ falls of rapidly. As a result, in matter dominated era only, the potential is reduced by a factor of $10^{-35}$. This might solve the cosmological constant problem as well without (possibly) requiring fine tuning, since it works within a range of parametric values.

\end{document}